\documentclass{article} 
\usepackage{amssymb}
\usepackage{amsmath}

\begin{document}
\title{How to Test for Diagonalizability: The Discretized \textsf{PT}-Invariant Square-Well Potential}
\author{Stefan Weigert
\\$\langle \mbox{Hu} |\mbox{MP} \rangle$ - Hull Mathematical Physics\\
          Department of Mathematics, University of Hull\\
          UK-Hull HU6 7RX}
%
%\headauthor{Stefan Weigert} \headtitle{How to Test for Diagonalizability} \lastevenhead{Stefan Weigert: How to Test for Diagonalizability}
%
%\pacs{03.67.-w} \keywords{$PT$-symmetry, diagonalizability,
%discretized square-well potential}
%%%%%%%%%%%%%% FOR EDITORIAL USE ONLY!!! %%%%%%%%%%%%%%%
%\refnum{A}%\total{}\type{}
%\daterec{XXX}    %;\\ final version }
%\issuenumber{0}  \year{2005}
%\setcounter{page}{1}
%\firstpage{1}
%\lastpage{000}
%\makefirsttitle
%%%%%%%%%%%%%%%%%%%%%%%%%%%%%%%%%%%%%%%%%%%%%%%%%%%%%%%%
\maketitle
%Page headings:

\begin{abstract}
Given a non-hermitean matrix \textsf{M}, the structure of its \emph{minimal}
polynomial encodes whether \textsf{M} is diagonalizable or not. This
note will explain how to determine the minimal polynomial of a matrix
without going through its \emph{characteristic} polynomial. The approach
is applied to a quantum mechanical particle moving in a square well
under the influence of a piece-wise constant \textsf{PT}-symmetric
potential. Upon discretizing the configuration space, the system is
decribed by a matrix of dimension three. It turns out not to be diagonalizable
for a critical strength of the interaction, also indicated by the
transition of two real into a pair of complex energy eigenvalues.
The systems develops a \emph{three-fold} degenerate eigenvalue, and
\emph{two} of the three eigenfunctions disappear at this exceptional
point, giving a difference between the \emph{algebraic} and \emph{geometric} multiplicity of the eigenvalue equal to two. 
\end{abstract}

%\pacs{03.67.-w}

\maketitle

\section{Introduction\label{sec:Introduction}}

Genuinely \textsf{PT}-invariant operators may or may not possess a
complete set of eigenstates. In other words, \textsf{PT}-invariance
of a matrix $\mathsf{M}$ is compatible with the presence of (non-trivial) Jordan blocks while hermiticity is not. When considering
a familiy of \textsf{PT}-invariant operators depending on a parameter,
their spectra often change qualitatively if one passes through an
\emph{exceptional} point \cite{kato84} where diagonalizability breaks
down. It is thus important to be able to either check whether a given
matrix is diagonalizable, or to locate exceptional points when presented
with a continuous family of matrices. 

The purpose of this note is to describe a method which allows one
to identify exceptional points of finite-dimensional non-hermitean
matrices by means of an algorithm. It is different from the method
outlined in \cite{abate97,weigert05a} as it directly aims at the
\emph{minimal} polynomial containing the relevant information about
(non-) diagonalizability. While being more transparent in the first
place, it also requires no knowledge of the \emph{characteristic}
polynomial of the given matrix. 

The presentation to follow is problem-based: the algorithm will be
developed while studying a specific example, the discretized \textsf{PT}-invariant
square well. This physical system is introduced in Section \ref{sec: discretized-PT-symmetric square well},
and it is subjected to the test for diagonalizability in the subsequent
section. The results will be discussed in Section \ref{sec:Summary-and-Discussion}
and some open questions will be addressed.

\section{The discretized PT-symmetric square well\label{sec: discretized-PT-symmetric square well}}

Consider a quantum particle in a one-dimensional box of length $4L$
subjected to a piece-wise constant \textsf{PT}-symmetric potential,

\begin{equation}
V(x)=\left\{ \begin{array}{cc}
\textrm{$-iZ,$} & -2L<x<0,\\
\quad\:0, & x=0\:,\\
\quad iZ, & \quad0<x<2L,\end{array}\right.\qquad Z\in\mathbb{R}\,,\label{eq: PT-potential}\end{equation}
which has proved a useful testbed for the discussion of \textsf{PT}-symmetric
systems. Its eigenvalues are given as the zeros of a transcendental
equation \cite{znojil01plus}, and, for each value of $Z$, the lowest
two real eigenvalues are known to coalesce and then disappear jointly
at critical values of the parameter. 

Let us introduce a toy-version of this system by discretizing its
configuration space. This strategy has been applied successfully to
decribe tunneling phenomena in a driven double-well potential in terms
of a three-state model \cite{schatzer+98}. Effectively, this technique
corresponds to turning Feynman's ``derivation'' of Schr\o"dinger's
equation from a discrete lattice \cite{feynman65} upside down. Explicitly,
the continuous set of points of configuration space with labels $-2L\leq x\leq2L$,
are replaced by \emph{five} equidistant points at $0,\pm L$, and
$\pm2L$. The wave function is allowed to take nonzero values only
at these points, so it will be a vector with five components at most.
However, the hard walls of the square well at $\pm L$ force the wave
function to vanish there, leaving us with only three non-zero components,
$\psi\rightarrow\psi_{k}=\psi(kL),k=0,\pm1$. The potential energy
defined in (\ref{eq: PT-potential}) turns into a diagonal matrix,
$V(x)\rightarrow\mathsf{V=}\mbox{{diag}}(-iZ,0,iZ)$. The operator
for the kinetic energy follows from replacing $\partial^{2}\psi(x)/\partial x^{2}\rightarrow(\psi_{k+1}-2\psi_{k}+\psi_{k-1})/L^{2}$.
Putting all this together, the Hamiltonian operator of the discrete
version of this system reads\begin{equation}
\mathsf{H_{0}}\simeq2\mathsf{E}-\mathsf{H}\:,\quad\mbox{where}\quad\mathsf{H}=\left(\begin{array}{ccc}
i\xi & 1 & 0\\
1 & 0 & 1\\
0 & 1 & -i\xi\end{array}\right),\quad\xi=Z/\eta\:;\label{eq:3by3 hamiltonian}\end{equation}
 here $\mathsf{E}$ is the $(3\times3)$ identity matrix, and an overall
factor $\eta=\hbar^{2}/2mL^{2}$ has been dropped. The matrix $\mathsf{H}_{0}$
inherits $\mathsf{PT}$-invariance from the square well: the matrix
$\mathsf{H}$, and hence $\mathsf{H}_{0}$, is invariant under the
combined action of parity \textsf{P}, represented by a matrix with
unit entries equal to one along its minor diagonal and zero elsewhere,
and \textsf{T,} effecting complex conjugation. In the next Section,
the diagonalizability of $\mathsf{H}$, the nontrivial part of the
Hamiltonian $\mathsf{H}_{0}$, will be studied.

\section{Diagonalizability of the PT-symmetric square well\label{sec: Square well diagonalizability}}

The \emph{minimal} polynomial \emph{}$m_{\mathsf{M}}$ of a matrix $\mathsf{M}$ is defined (see \cite{lancaster+85}, for example) as
the polynomial of least degree in $\mathsf{M}$ which annihilates
$\mathsf{M}$, that is, $m_{\mathsf{M}}(\mathsf{M})=0$. This polynomial
is unique if the coefficient of its highest power is taken to be one:
the minimal polynomial is \emph{monic.} Since any matrix $\mathsf{M}$
of size $N$, say, is annihilated by its own characteristic \emph{polynomial,}
$p_{\mathsf{M}}(\mathsf{M})=0$\emph{,} the degree of the minimal
polynomial does not exceed $N$. Once $m_{\mathsf{M}}$ has been found,
one needs to determine whether it has only \emph{single} roots, i.e.
whether\begin{equation}
m_{\mathsf{M}}(\lambda)=\prod_{\nu=1}^{\nu_{0}(\leq N)}(\lambda-M_{\nu})\,,\quad\mbox{all }\, M_{\nu}\mbox{ distinct}\,.\label{genminpolydiag}\end{equation}
 holds. If it does, the matrix $\mathsf{M}$ \emph{is} diagonalizable
- otherwise, it is \emph{not} diagonalizable since multiple roots
of $m_{\mathsf{M}}$ indicate the presence of Jordan blocks
larger than one. 

The procedure to determine the minimal polynomial of $\mathsf{M}$,
as outlined in \cite{weigert05a}, invokes the characteristic polynomial
of $\mathsf{M}$ and repeated applications of the Euclidean algorithm
generalized to polynomials. The method presented below, taken from
\cite{horn+85}, aims at directly constructing the minimal polynomial.
The fundamental observation is that matrices of size $N$ constitute
a vector space of dimension $N^{2}$. This is seen immediatly by setting
up a one-to-one correspondence between matrices of size 3 and vectors
of length $9\equiv3^{2}$, for example, simply by rearranging the
elements $\mathsf{M}_{jk}$ of each $\mathsf{M}$ systematically according
to\begin{equation}
\mathsf{M}\Leftrightarrow(\mathsf{M}_{11},\mathsf{M}_{12},\mathsf{M}_{13};\mathsf{M}_{21},\mathsf{M}_{22},\mathsf{M}_{23};\mathsf{M}_{31},\mathsf{M}_{32},\mathsf{M}_{33})^{T}.\label{eq: correspondence}\end{equation}
 In view of this correspondence, the theorem by Cayley-Hamilton -
every matrix satisfies its own characteristic equation, $p_{\mathsf{M}}(\mathsf{M})=0$,
- turns into a statement about linear \emph{dependence} of the \emph{$(N+1)$
vectors} $\mathsf{E\equiv M}^{0},\mathsf{M},\mathsf{M}^{2},\ldots,\mathsf{M}^{N}$.
In order to find the minimal polynomial of a matrix $\mathsf{M}$
one thus simply calculates $\mathsf{M}^{n},n=1\ldots N,$ and then
determines successively whether the vectors $\mathsf{E}$ and $\mathsf{M}$
are linearly independent; if not, one adds $\mathsf{M}^{2}$ and asks
the same question; etc. Proceeding in this way, one is obviously able
to identify linear dependence among the vectors $\mathsf{M}^{n}$
containing only the smallest powers necessary. This, however, comes
down to the definition of the minimal polynomial of \textsf{M}. Gram-Schmidt
orthonormalization effectively provides a systematic test for linear
dependence among the first $k$ elements of $\mathsf{M}^{n},n=0\ldots N$. 

Applying these ideas explicitly to the matrix $\mathsf{H}$ in (\ref{eq:3by3 hamiltonian})
leads to \begin{eqnarray}
\mathsf{E} \, \,  & \Leftrightarrow & (1,0,0;0,1;0;0,0,1)\:,\\
\mathsf{H} \, \,  & \Leftrightarrow & (i\xi,1,0;1,0,1;0,1,-i\xi)\:,\\
\mathsf{H}^{2} & \Leftrightarrow & (1-\xi^{2},-i\xi,1;-i\xi,2,i\xi;1,i\xi,1-\xi^{2})\:,\\
\mathsf{H}^{3} & \Leftrightarrow & (2-\xi^{2})\:(i\xi,1,0;1,0,1;0,1,-i\xi)\:\equiv(2-\xi^{2})\mathsf{H\:}.\end{eqnarray}
It is easy to see that neither the first two nor the first three vectors in this sequence are linearly dependent. Consequently, there must be a relation expressing $\mathsf{H}^{3}$ in terms of the others, and indeed, the minimal polynomial follows immediately from 
\begin{equation}
(2-\xi^{2})\mathsf{H}-\mathsf{H}^{3}=0\quad\Rightarrow\quad m_{\mathsf{H}}(\lambda)=\lambda^{3}+(\xi^{2}-2)\lambda\:.\label{eq: minipolyH}\end{equation}
 In addition, the characteristic polynomial of $\mathsf{H}$ must
coincide with $m_{\mathsf{H}}$ since it there is only one monic polynomial of third degree annihilating ${\mathsf{H}}$. 

If one is not able to factor the resulting minimal polynomial, one
needs to check whether the minimal polynomial and its derivative $m^{\prime}\mathsf{_{H}}(\lambda)$
have a common factor which can be achieved by applying the Euclidean
algorithm to this pair (cf. \cite{abate97,weigert05a}). In this present
case, this amounts to writing $m\mathsf{_{H}}(\lambda)=(\lambda-\alpha)m_{\mathsf{H}}^{\prime}(\lambda)/3+R_{1}(\lambda)$,
implying that $\alpha=0$ and $R_{1}(\lambda)=(2/3)(\xi^{2}-2)(\lambda+1/2)$.
Two different cases arise: if $\xi\neq2$, one finds that the only
common factor of the minimal polynomial and the derivative is equal
to one - thus, the minimal polynomial is of the form (\ref{genminpolydiag})
and the matrix $\mathsf{H}$ must have three different eigenvalues
making it diagonalizable. If $\xi^{2}=2,$ the algorithm immediately
stops and thus identifies $\lambda^{2}$ as the highest common factor
of $m_{\mathsf{H}}(\lambda)$ and its derivative. Consequently, the
minimal polynomial has a three-fold root $\lambda(=0)$, indicating
that $\mathsf{H}$ is not diagonalizable.

\section{Discussion and Outlook\label{sec:Summary-and-Discussion}}

The properties of $\mathsf{H}$ at the exceptional points defined
by $\xi_{\pm}=\pm\sqrt{{2}}$ deserve a brief discussion. It is not
difficult to see that the \emph{geometric} multiplicity of the eigenvalue
0 is one at the exceptional points ($\mathsf{H}$ has only one non-zero
eigenvector) while its \emph{algebraic} multiplicity equals three
(zero is a triple root of $m_{\mathsf{H}}$). Contrary to previously
studied cases, \emph{three} eigenvalues coalesce for $\xi\rightarrow\xi_{\pm}$.
For all values of $\xi$, the vector $(1,-i\xi,-1)$ is an eigenstate
of $\mathsf{H}(\xi)$ with eigenvalue 0, and it does not exhibit any
particular behaviour at the critical values $\xi_{\pm}$. Therefore,
it seems reasonable to say that the eigenstates associated with the
two $\lambda$-dependent eigenvalues disappear at the exceptional
point. It is not obvious from a physical point of view why this scenario
is preferred over the familiar situation of just one disappearing
eigenstate. 

The natural question to ask now is whether one can expect algorithmic
tests for diagonalizability to exist for a quantum system living in
a Hilbert space accomodating a countable infinity of states. As one
needs to potentially perform an infinite number of steps, the idea
of a useful algorithm gets somewhat blurred. Nevertheless, it is likely
that one can search for systematic properties of finite-dimensional
approximations which, hopefully, behave smoothly in the limit of infinite
dimension. 

\end{document}